\documentclass[conference]{IEEEtran}
\IEEEoverridecommandlockouts
\usepackage{cite}
\usepackage[utf8]{inputenc}
\usepackage[T1]{fontenc}
\usepackage{amsmath,amssymb,amsfonts}
\usepackage{algorithmic}
\usepackage{graphicx}
\usepackage{textcomp}
\usepackage{xcolor}
\usepackage{array}
\usepackage{algorithm2e}
\usepackage{floatrow}
\usepackage{caption}
\usepackage{amsthm} % 用于定理环境，必须在 hyperref 之前
\usepackage{geometry}
\usepackage{subcaption}
\usepackage{setspace}
\usepackage[absolute,overlay]{textpos} % 只加载一次

\usepackage[hypertexnames=false]{hyperref} % 删除重复的 hyperref 加载

\def\BibTeX{{\rm B\kern-.05em{\sc i\kern-.025em b}\kern-.08em
    T\kern-.1667em\lower.7ex\hbox{E}\kern-.125emX}}
\floatstyle{plaintop} 
\restylefloat{table}
\makeatletter
\def\ps@IEEEtitlepagestyle{%
  \def\@oddhead{%
    \parbox{\textwidth}{\centering
      \normalsize
      L.~Xue, S.~Sun, H.~Yan,
      ``Spatial Correlation and Degrees of Freedom in Arched HMIMO Arrays: A Closed-Form Analysis,''
      in \textit{2025 IEEE 101st Vehicular Technology Conference (VTC WKSP)}, Oslo, Norway, 2025.}%
  }%
  \def\@evenhead{\@oddhead}%
  \def\@oddfoot{}%
  \def\@evenfoot{}%
}
\makeatother
\begin{document}

\title{Spatial Correlation and Degrees of Freedom in Arched HMIMO Arrays: A Closed-Form Analysis}

\author{
\IEEEauthorblockN{Liuxun Xue\IEEEauthorrefmark{1}, Shu Sun\IEEEauthorrefmark{1}, and Hangsong Yan\IEEEauthorrefmark{2}}
\IEEEauthorblockA{\IEEEauthorrefmark{1} School of Information Science and Electronic Engineering,
Shanghai Jiao Tong University, Shanghai 200240, China\\}
\IEEEauthorblockA{\IEEEauthorrefmark{2}Hangzhou Institute of Technology, Xidian University, Hangzhou, Zhejiang 311231, China\\
Corresponding author: Shu Sun (Email: shusun@sjtu.edu.cn)}
}

\maketitle

\begin{abstract}
This paper presents a closed-form analysis of spatial
correlation and degrees of freedom (DoF) for arched holographic
multiple-input multiple-output (HMIMO) arrays, which can
be viewed as a special form of fluid antenna systems (FAS)
when their geometry is fluidically adaptable. Unlike traditional
planar configurations, practical HMIMO surfaces may exhibit
curvature, significantly influencing their spatial characteristics
and performance. We derive exact correlation expressions for
both arched uniform linear arrays and arched uniform
rectangular arrays, capturing curvature effects under far
field propagation. Our results reveal that isotropic scattering results in DoF being dominated by the maximum span of the HMIMO array, such that shape effects are weakened, and bending does not significantly reduce the available spatial DoF. Numerical simulations validate the accuracy of the closed-form formulas and demonstrate the robustness of DoF against curvature variations, supporting flexible array designs. These findings offer fundamental insights into geometry-aware optimization for next-generation HMIMO/FAS systems and pave the way for practical implementations of curved HMIMO arrays.
\end{abstract}

\let\thefootnote\relax\footnotetext{
This work was supported in part by the National Natural Science Foundation of China under Grant 62271310 and Grant 62431014, and in part by the Fundamental Research Funds for the Central Universities of China.}

\section{Introduction}

Driven by the ambitious performance targets of 6G and beyond wireless networks, a new wave of \emph{reconfigurable antenna} technologies has emerged to enable flexible and adaptive communications. Among these, \emph{fluid antenna systems} (FAS) have gained significant attention~\cite{FAS1}. FAS broadly encompasses software-controlled fluidic, dielectric, or conductive structures, such as mechanical liquid antennas, RF pixel antennas, movable antennas, or flexible metasurfaces, whose shape, size, orientation, or position can be reconfigured in real time~\cite{FAS1,FAS2}. This dynamic geometry fully exploits spatial variations, enhancing reliability and throughput. In parallel, \emph{holographic multiple-input multiple-output} (HMIMO), which can be regarded as a special category of FAS~\cite{FAS2}, has been envisioned as a key enabler in 6G wireless networks~\cite{Ref_huang2020holographic}.  
% HMIMO focuses on dense antenna elements on nearly continuous surfaces for high coverage, capacity, and energy efficiency. Though distinct in implementation, HMIMO, reconfigurable intelligent surfaces (RIS), and FAS share a principle of reconfigurability: HMIMO/RIS adjust dense or tunable elements, while FAS reconfigures by physically reshaping radiating apertures. Thus, FAS can be seen as a fluidic HMIMO, adding true geometric flexibility.

In real-world scenarios, surfaces where HMIMO arrays are situated on are often not perfectly planar (e.g., walls, pillars, vehicles, or unmanned aerial vehicles). Such arched or conformal geometries \cite{Ref_ConformalRadar, Ref_wu2023enabling} affect channel rank, degrees of freedom (DoF), and capacity. When fluidic or mechanical adaptation is also possible, these arched HMIMO arrays become a specialized form of FAS. A key performance indicator in such systems is the spatial DoF~\cite{Ref_sun2025differentiate, Ref_PizzoMarzetta2020}. Higher DoF implies a higher spatial rank, boosting spectral efficiency or resilience to fading or interference. Yet, most HMIMO studies assume planar arrays \cite{Ref_PizzoMarzetta2020}, leaving it unclear how curvature affects DoF and capacity at scale.

Under planar assumptions, DoF typically grows with aperture size in the far field. While some works address linear, planar, volumetric \cite{Ref_sun2021small, Ref_SunTao2022, Ref_PizzoMarzetta2020} or spherical arrays \cite{Ref_poon2005degrees}, direct extensions of planar-to-curved arrays remain limited. Conformal-antenna research \cite{Ref_ConformalRadar} outlines beam patterns and efficiency but lacks a thorough treatment of spatial correlation and DoF under curvature. Hence, a more comprehensive framework is needed to realize the full potential of arched or fluidic HMIMO surfaces for 6G.

In this paper, we introduce a closed-form analysis of spatial correlation and DoF for arched HMIMO arrays, a specialized form of FAS. Our main contributions are as follows:
 \begin{itemize}
   \item We derive exact spatial correlation expressions for an \emph{arched uniform linear array} (ULA) under a half-space isotropic scattering environment. The result extend classical ULA analysis, quantifying how curvature alters the correlation matrix and impacts the system's ability to support independent data streams.
   \item We then generalize the derivation to a two-dimensional arched uniform rectangular array (URA). A closed-form correlation solution is given, covering both planar and semi-cylindrical extremes.
   \item Analysis of the eigenvalue spectra reveals that both the arched ULAs and arched URAs exhibit only a slight decrease in DoF with increasing curvature, while the overall DoF remains relatively stable.
\end{itemize}
 
Results for the arched ULAs and URAs offer key insights for designing practical HMIMO arrays under curved geometries. Despite variations in curvature, our analysis indicates that its impact on DoF remains limited under far field conditions and isotropic scattering environments, thereby offering greater design flexibility for the implementation of HMIMO/FAS systems across diverse geometries

% In this paper, we propose the closed-form analysis of spatial correlation and degrees of freedom (DoF) for \emph{arched HMIMO arrays} （A specialized example of fluid antenna systems）. Our contributions can be summarized as follows:
% \begin{itemize}
%     \item We derive exact spatial correlation expressions for an \emph{arched uniform linear array} under a half-space isotropic scattering environment and also derive the expression for the average DoF limit of the arched ULA. These results extend classical ULA analysis, quantifying precisely how curvature alters the correlation matrix and impacts the system's ability to support independent data streams.
%     \item We then generalize our derivations to a two-dimensional \emph{arched uniform rectangular array}, capturing the interplay between curvature along one dimension and array extension along the orthogonal dimension. A closed-form solution for the correlation entries is also provided, encompassing both planar and semi-cylindrical limits.
%     \item By examining the eigenvalue spectra of the correlation matrices, we show that while the arched ULA exhibits monotonically decreasing DoF with curvature, the arched URA displays a distinctive non-monotonic trend.
% \end{itemize}

\begin{figure*}[htbp]
\centerline{\includegraphics[width=0.7\linewidth]{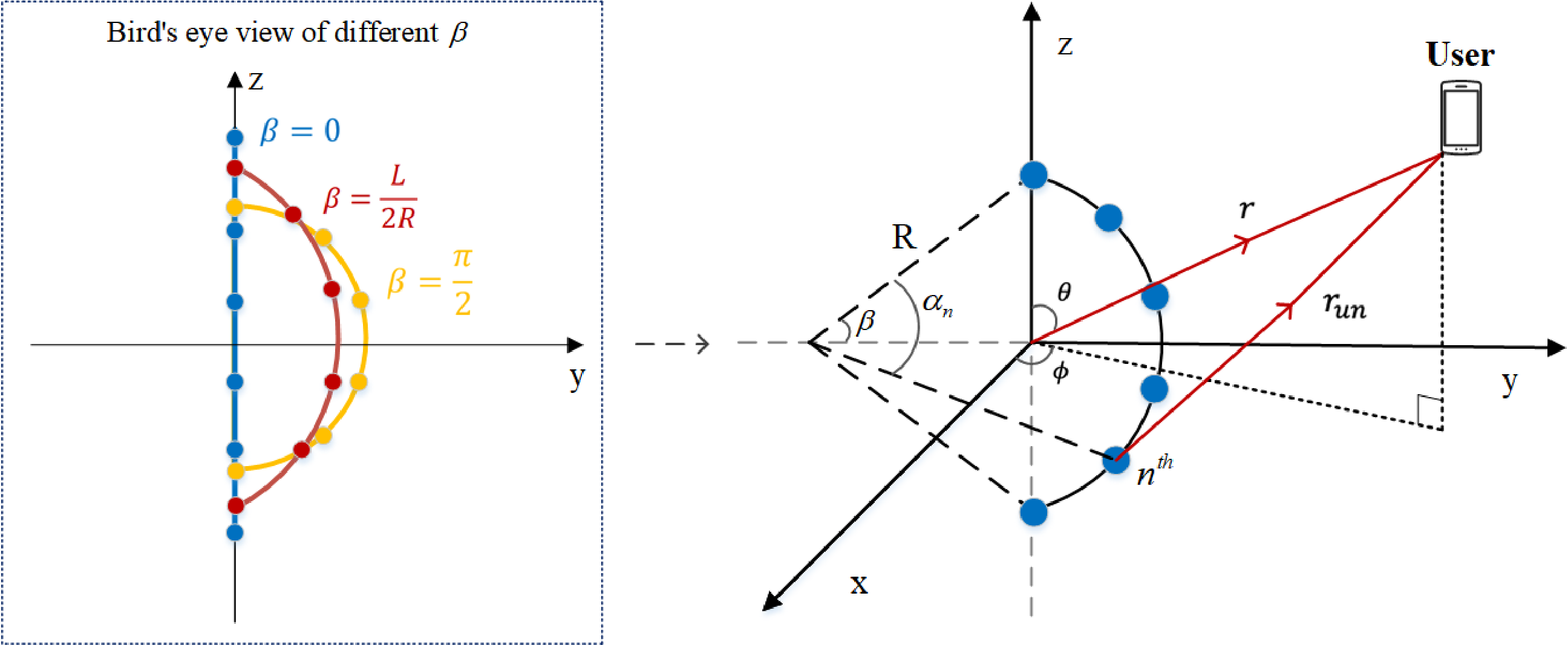}}
\captionsetup{font=footnotesize, labelsep=period}
\caption{Geometry and bird's eye view of the arched ULA in far field.}
\label{ULA}
\end{figure*}

% Results for both the arched ULA and URA provide valuable insights for designing practical HMIMO arrays, where curved geometries are likely to arise. Our analysis can aid in the development of more efficient and flexible HMIMO systems, facilitating better spatial diversity and resolution while addressing the challenges posed by curved array configurations.

% The paper is organized as follows. In Section~II, we introduce the geometry and far field steering vectors of arched ULA and URA. Section~III provides closed-form derivations of the spatial correlation and a DoF analysis for both array types. Numerical results are presented in Section~IV, followed by concluding remarks and future directions in Section~V.

\section{System Model}
\label{2}
We consider a downlink scenario where a base station employs an arc-shaped HMIMO array, encompassing both arched ULA and URA configurations, to serve single-antenna users in three-dimensional (3D) space, as illustrated in Fig. 1. 

\subsection{Arched ULA Geometry}
\label{ArchedULAGeometry}
 
Consider a holographic arched ULA with total arc length \(L\) and curvature radius \(R\). The array is positioned vertically in the YZ-plane, as illustrated in Fig.~\ref{ULA}. Let \(N\) denote the number of antenna elements in the arched ULA, which are evenly spaced along the arc, with element spacing \(d_{yz}\). A single-antenna user is assumed to be located at \(\bigl(r\sin(\theta)\cos(\phi),\;r\sin(\theta)\sin(\phi),\;r\cos(\theta)\bigr)\), where \(r\) is the radial distance between the origin and the user, \(\theta\) is the zenith angle, and \(\phi\) is the azimuth angle.  Moreover, the coordinates of the \(n\)-th element can be written as 
\begin{align}
(0,\;R\cos(\beta-\alpha_n) - R\cos(\beta),\;R\sin(\beta-\alpha_n)),
\end{align}
where \(\beta = \dfrac{L}{2R} \in \bigl[0,\tfrac{\pi}{2}\bigr]\) is the bending angle (i.e., half of the central angle corresponding to the total arc length \(L\)), and \(\alpha_n\) denotes the central angle corresponding to the \(n\)-th element, defined as \(\alpha_n \;=\; \frac{(n-1)\,L}{(N-1)\, R}, 
\quad n=1,\dots,N.\)
 
Thus, the exact distance between the user and the \(n\)-th element, denoted by \(r_{un}\), is given by  
\begin{equation}
r_{un} = \sqrt{
\begin{aligned}
&\bigl[r\,\sin(\theta)\cos(\phi)\bigr]^2
\\+&\Bigl[r\,\sin(\theta)\sin(\phi) - R\,\cos(\beta-\alpha_n) 
    + R\,\cos(\beta)\Bigr]^2
 \\+&\bigl[r\,\cos(\theta) 
         - R\,\sin(\beta-\alpha_n)\bigr]^2. 
         \end{aligned}
         } 
\end{equation}
In classical MIMO systems that assume a far field plane-wave model (i.e., \(r \gg L\)). Only the first-order term linearly dependent on the arc radius \(R\) is retained from (2) to capture the dominant inter-element phase differences, thereby yielding the approximate expression for \(r_{un}\)
\begin{equation}
{\footnotesize
\begin{split}
r_{un}
&\approx 
r - R \left[ \sin (\theta) \sin (\phi) \left( \cos(\beta - \alpha_n) - \cos \beta \right) + \cos (\theta) \sin(\beta - \alpha_n) \right].
\end{split}
}
\label{eq:d_un_approx}
\end{equation}
% This simplification captures the arc geometry’s effect on the propagation distance while remaining tractable for analytical design. 
Consequently, the array steering vector for the holographic arched ULA can be written as
\begin{equation}
{\footnotesize
\begin{split}
& \mathbf{a}(\theta, \phi) \\
 = & \frac{1}{\sqrt{N}} \biggl[ 
  e^{j \frac{2 \pi R}{\lambda} \left[ \sin (\theta) \sin (\phi) \left( \cos(\beta - \alpha_1) - \cos (\beta) \right) + \cos (\theta) \sin(\beta - \alpha_1) \right]}, \\
  & \ldots, 
  e^{j \frac{2 \pi R}{\lambda} \left[ \sin (\theta) \sin (\phi) \left( \cos(\beta - \alpha_n) - \cos (\beta) \right) + \cos (\theta) \sin(\beta - \alpha_n) \right]} \biggr]^T.
\end{split}
}
\label{eq:archedULA_steer2}
\end{equation}
where \(\lambda\) is the carrier wavelength. 

\subsection{Arched URA Geometry}
\label{ArchedURAGeometry}
In practical applications, uniform planar arrays are preferred over ULAs. Likewise, the arched ULA can be extended to an arched URA, which improves coverage in both the azimuth and elevation planes. The array geometry is shown in Fig. 2. 
% Thus, Consider a holographic arched URA, 
From a geometric perspective, the arched URA can be viewed as a collection of \(M\) concentric arched ULA segments, each placed along the \(\mathrm{X}\)-axis at \(x = m\,d_x\) for \(m = 1,\dots, M\), with a uniform element spacing of \(d_x\).

In this section, we focus on an $M$-layer arched URA structure, considering an arched URA with element $(m,n)$ placed at
% In this section, we focus on a $M$-layer 3D arched URA array structure, where each layer is an arched ULA array as analyzed in Section II A, each placed along the \(x\)-axis at \(x = m\,d_x\) for \(m = 0,1,\dots,M\), with a uniform element spacing of \(d_x\). Consider an arched URA with element $(m,n)$ coordinates placed at
\begin{equation}
\mathbf{r}_{m,n} 
= \bigl(md_x,\; R\cos(\beta - \psi_n)-R\cos(\beta),\;R\sin(\beta - \psi_n)\bigr),
\label{eq:element_coord}
\end{equation}

\begin{figure*}[htbp]
\centerline{\includegraphics[width=.9\linewidth]{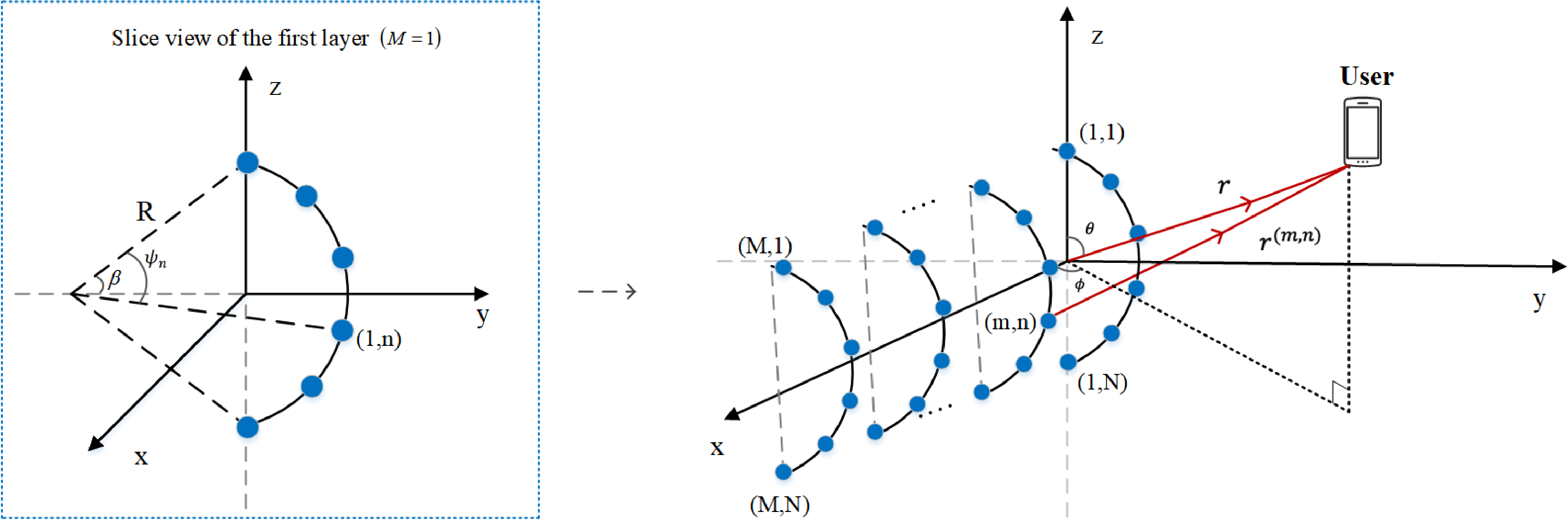}}
\captionsetup{font=footnotesize, labelsep=period}
\caption{Geometry and the first layer slice view of the arched URA in 3D space.}
\label{URA}
\end{figure*}

\noindent where $R$ and $L$ denote the radius of curvature and length of each arched ULA segment, respectively, $\beta = \frac{L}{2R} \in \bigl[0,\tfrac{\pi}{2}\bigr]$ denotes the arc opening angle, and $\psi_n =\frac{(n-1)\,L}{(N-1)\,R}\in[0,\pi]$ parameterizes the $n$-th element in each arched ULA. Assume a far field source at distance \(r\gg R, md_x\), described by spherical angles $(\theta,\phi)$. Following a similar process in Section~\ref{ArchedULAGeometry}, the 3D distance between the user and the \((m,n)\)-th element is written as
\begin{equation}
r^{(m,n)} = \sqrt{
    \begin{aligned}
    &[r\sin(\theta)\cos(\phi) - md_x]^2 \\
    +&\left[r\sin(\theta)\sin(\phi) - R\cos(\beta-\psi_n) + R\cos(\beta)\right]^2 \\
    +&\left[r\cos(\theta) - R\sin(\beta-\psi_n)\right]^2.
    \end{aligned}
}
\end{equation}
% Under far field assumption (\(r\gg R, md_x\)), second-order small quantity can usually be ignored when approximating to the first order, so this simplifies to:
Under the far field assumption (\(r\gg R, md_x\)), this simplifies to
\begin{equation}
\begin{aligned}
  r^{(m,n)} &\approx r - \Delta_{m,n}, \\
  \Delta_{m,n} &= m\,d_x\,\sin(\theta)\,\cos(\phi) 
  \;\\&+\; R\Bigl[\sin(\theta)\,\sin(\phi)\,\bigl(\cos(\beta-\psi_n) - \cos(\beta)\bigr) 
    \;\\&+\; \cos(\theta)\,\sin(\beta-\psi_n)\Bigr].
\end{aligned}
\end{equation}
Thus, the steering vector of the arched URA is given by
\begin{equation}
\scalebox{1}{$
\mathbf{a}_\mathrm{URA}(\theta,\phi)
= \frac{1}{\sqrt{N}}
\Bigl[
e^{j \tfrac{2\pi}{\lambda}\,\Delta_{m_1,n_1}},
\,\dots,\,
e^{j \tfrac{2\pi}{\lambda}\,\Delta_{m_N,n_N}}
\Bigr]^{T}.
\label{eq:archedURA_difference}
$}
\end{equation}

\section{Spatial Correlation: Closed-Form Solutions and Degrees of Freedom Analysis}

In this section, we derive closed-form spatial correlation functions and analyze the average DoF limit of the arched HMIMO arrays. We begin with the arched ULA due to its simpler geometry, then extend our findings to the more general arched URA configuration.

\subsection{Closed-Form Analysis for Arched ULA}

\label{subsec:ULA_analysis}  
% In this subsection, we conduct a more detailed analysis of spatial correlation of the considered arched ULA.
% We rigorously derive the spatial correlation for the arched ULA. 
% under half-space isotropic scattering. 
% This closed-form model quantifies curvature effects on spatial statistics, a critical advancement for HMIMO system design.  
For a half-space isotropic scattering environment, the spatial scattering function is~\cite{Ref_SunTao2022}
\begin{equation}
  S(\theta,\phi) = \frac{\sin(\theta)}{2\pi}, \theta \in \bigl[0,\pi\bigr], \phi \in \bigl[0,\pi\bigr].  
\end{equation}
The spatial correlation matrix is defined as
\begin{equation}
\mathbf{R}_0 = \int_0^{\pi} \int_0^{\pi} S(\phi, \theta) \textbf{a}(\phi, \theta) \textbf{a}^H(\phi, \theta) \, d\phi \, d\theta.  
\end{equation}
The correlation matrix entry hence becomes
\begin{equation}
\begin{aligned}
[\mathbf{R}]_{m,n} 
&= \int_0^\pi \int_0^\pi \frac{\sin(\theta)}{2\pi} e^{j\Delta_{m,n}(\theta, \phi)} d\theta d\phi,
% \\&= \frac{1}{2} \int_0^\pi \sin(\theta) \, e^{j\Delta_{m,n}(\theta)} d\theta,
\end{aligned}
\label{eq:ULA_corr_integral}
% \int_{0}^{\pi}\int_{0}^{\pi} 
% \frac{\sin(\theta)}{2\pi} 
% \, e^{\,j \tfrac{2\pi R}{\lambda} 
% \Bigl[\sin\bigl(\theta+\beta-\alpha_{m}\bigr) 
% - \sin\bigl(\theta+\beta-\alpha_{n}\bigr)\Bigr]} 
% \,d\phi\,d\theta.
\end{equation}
where $\Delta_{m,n}(\theta)$ represents the phase difference of elements \(m\) and \(n\) on the arched ULA under far field conditions, which is written as
\begin{align}
&\Delta_{m,n} (\theta, \phi) \nonumber \\
 = & \frac{2\pi R}{\lambda} \left[ \sin (\theta) \sin (\phi) \left( \cos(\beta - \alpha_n) - \cos(\beta - \alpha_m) \right) \right. \nonumber \\ 
 & \left. + \cos (\theta) \left( \sin(\beta - \alpha_n) - \sin(\beta - \alpha_m) \right) \right],
\label{ULA_difference}
\end{align}
%\begin{equation}
%{\small
%\begin{aligned}
%\Delta_{m,n}(\theta, \phi) &= \frac{2\pi R}{\lambda} \left[ \sin (\theta) \sin (\phi) \left( \cos(\beta - \alpha_n) - \cos(\beta - \alpha_m) \right) 
%\\& + \cos (\theta) \left( \sin(\beta - \alpha_n) - \sin(\beta - \alpha_m) \right) \right],
%\label{ULA_difference}
%\end{aligned}
%}
%\end{equation}  
where $\alpha_m = \frac{(m-1)L}{(N-1)R}$.
\\
\subsubsection{Step-by-Step Derivation via Bessel Expansion} Simplify (12) using trigonometric identities, and the spatial correlation becomes
\begin{equation}
{\small
\begin{split}
[\mathbf{R}]_{m,n} = \int_0^\pi \int_0^\pi \frac{\sin (\theta)}{2 \pi} e^{-j b \left[ \sin (\theta) \sin (\phi) \sin (c) - \cos (\theta) \cos (c) \right]} \, d\phi \, d\theta,      
\end{split}
}
\end{equation}
where $b = \frac{4 \pi R}{\lambda} \sin \Delta \alpha = \frac{4 \pi R}{\lambda} \sin \left( \frac{\alpha_n - \alpha_m}{2} \right)$, $c = \beta - \frac{\alpha_m + \alpha_n}{2}$. 
Then, compute the azimuth integral
\begin{equation}
\begin{aligned}
I_\phi &= \int_0^\pi e^{-j b \left( \sin (\theta) \sin (\phi) \sin (c) - \cos (\theta) \cos (c) \right)} \, d\phi,
\\&= e^{j b \cos (\theta) \cos (c)} \int_0^\pi e^{-j b \sin (\theta) \sin (\phi) \sin (c)} \, d\phi. 
\end{aligned}
\end{equation}
Let \( z = b \sin (\theta) \sin (c) \), and apply the Jacobi-Anger expansion~\cite{Ref_bowman2012introduction}
\begin{equation}
\begin{aligned}
e^{-j z \sin (\phi)} = \sum_{k=-\infty}^\infty J_k(z) e^{-j k (\phi)},
\end{aligned}
\end{equation}

\begin{equation}
\int_0^\pi e^{-j z \sin (\phi)} \, d\phi = \sum_{k=-\infty}^\infty J_k(z) \int_0^\pi e^{-j k (\phi)} \, d\phi,    
\end{equation}

\begin{equation}
\int_0^\pi e^{-j k (\phi)} \, d\phi = \begin{cases} 
\pi, & k = 0 \\
\frac{(-1)^k - 1}{-j k}, & k \neq 0 
\end{cases}.
\end{equation}
Thus,
\begin{equation}
\int_0^\pi e^{-j z \sin (\phi)} \, d\phi = \pi J_0(z) + \sum_{k=1,3,5,\ldots} \frac{2}{j k} J_k(z), 
\end{equation}
and the azimuth integral $I_\phi$ is given by
\begin{equation}
{\small
\begin{split}
I_\phi &= e^{j b \cos (\theta) \cos (c)} \\&\left[ \pi J_0(b \sin (\theta) \sin (c)) + \sum_{k=1,3,5,\ldots} \frac{2}{j k} J_k(b \sin (\theta) \sin(c)) \right].    
\end{split}
}
\end{equation}
Substitute into (11), accounting for the weighting factor \( \frac{1}{2 \pi} \)
\begin{equation}
{\footnotesize
\begin{split}
[\mathbf{R}]_{m,n} &= \int_0^\pi \sin (\theta) e^{j b \cos (\theta) \cos (c)} \frac{J_0(b \sin (\theta) \sin (c)) }{2} d\theta \\&+ \int_0^\pi \sin (\theta) e^{j b \cos (\theta) \cos (c)} \sum_{k=1,3,5,\ldots} \frac{J_k(b \sin (\theta) \sin (c))}{j k \pi}   d\theta.   
\end{split}
}
\end{equation}

\subsubsection{Analytical Evaluation of the Zero-Order Component}
Specifically, for the zero-order term ($k=0$), we have
\begin{equation}
[\mathbf{R}]_{m,n,\text{zero}} = \int_0^\pi \sin (\theta) e^{j b \cos (\theta) \cos (c)} \cdot \frac{J_0(b \sin (\theta) \sin (c))}{2}  \, d\theta.
\end{equation}
To evaluate this integral, we perform a change of variable by letting \( u = \cos (\theta) \), yielding
\begin{equation}
= \frac{1}{2} \int_{-1}^1 e^{j b u \cos (c)} J_0(b \sqrt{1 - u^2} \sin (c)) \, du.
\end{equation}
This integral matches a known form~\cite{Ref_bowman2012introduction}
\begin{equation}
\int_{-1}^1 J_0\bigl(a \sqrt{1-x^2}\bigr)\,e^{\,jax}\,dx 
\;=\;
\frac{2\,\sin\!\bigl(\sqrt{a^2 + b^2}\bigr)}{\sqrt{a^2 + b^2}}.
\end{equation}
Applying this identity with appropriate substitutions yields the closed-form zero-order term as
\begin{equation}
[\mathbf{R}]_{m,n,\text{zero}} = \text{sinc}\left(\frac{4 R}{\lambda} \sin \left( \frac{\alpha_n - \alpha_m}{2} \right)\right).
\end{equation}

\setcounter{footnote}{0}  % 重置脚注计数器
\def\thefootnote{\arabic{footnote}}  % 恢复数字脚注编号
\subsubsection{Negligibility of the Odd-Order Components}
% {Asymptotic Negligibility of the Odd-Order Components}
The odd-order term (odd $k$) is explicitly given by
\begin{equation}
{\normalsize
\begin{split}
[\mathbf{R}]_{m,n,\text{odd}} &= \int_0^\pi \sin (\theta) e^{j b \cos (\theta) \cos (c)}  \\& \left( \sum_{k=1,3,5,\ldots} \frac{1}{j k \pi} J_k(b \sin (\theta) \sin (c)) \right) d\theta.
\end{split}
}
\end{equation}
The resulting integral is complex and typically evaluated numerically, as the odd-order terms are oscillatory and lack a simple closed form. However, due to the rapid decay of $J_k(z)$ for large $k$ and the averaging effect over $\theta$, we hypothesize that their overall contribution is negligible.\footnote{To test the hypothesis that the odd-order terms are negligible, we conducted numerical simulations, which confirmed they are significantly smaller than the zero-order term, details are omitted due to space constraints.}
Thus, we obtain Theorem 1 below.
\\
\textbf{Theorem 1.} \textit{The closed-form spatial correlation for the arched ULA can be formulated as}
\begin{align}  
   [\mathbf{R}]_{m,n} = \text{sinc}\left(\frac{4 R}{\lambda} \sin \left( \frac{\alpha_n - \alpha_m}{2} \right)\right).  
   \label{eq:ULA_final}  
   \end{align}  
\\
\textit{Remark 1.}
Theorem 1 shows that for an arched ULA, spatial correlation depends on the angular separation between elements.

% From the explicit expression in Theorem 1, we deduce that the dominant terms \((k=0, \pm 1)\) govern the primary spatial correlation characteristics.

% \begin{figure*}[htbp]
% \centerline{\includegraphics[width=.9\linewidth]{URAAA.png}}
% \captionsetup{font=footnotesize, labelsep=period}
% \caption{Geometry and the first layer slice view of the arched URA in 3D space.}
% \label{URA}
% \end{figure*}

\subsection{Closed-Form Analysis for Arched URA}
Next, we extend the closed-form derivation to the arched URA. 
\subsubsection{Phase Difference}
Consider the array elements indexed by $(m,n)$ and $(m',n')$. As derived in Section~\ref{ArchedURAGeometry}, the phase difference from the planar wave characterized by angles \(\theta\) and \(\phi\) takes the form
\begin{align}
&\Delta_{m,n} - \Delta_{m',n'} \;
\\&=\; (m - m')\, d\, \sin(\theta) \cos(\phi) 
\nonumber \\
&+ R \sin(\theta) \sin(\phi) \Big[\cos(\beta - \psi_n) - \cos(\beta - \psi_{n'})\Big]
\nonumber \\
&+ R \cos(\theta) \Big[\sin(\beta - \psi_n) - \sin(\beta - \psi_{n'})\Big].
\label{eq:phasediff_original}
\end{align}
Then, using the trigonometric identities, \eqref{eq:phasediff_original} can be rewritten in the form
\begin{equation}
\Delta_{m,n} - \Delta_{m',n'} \;=\; A(\theta)\cos(\phi) + B(\theta)\sin(\phi) + C(\theta),
\end{equation}
where
\begin{align}
A(\theta) \;&=\; (m - m')\, d \,\sin(\theta),\\
B(\theta) \;&=\; -2R \,\sin(\theta) \,\sin\!\Bigl(\beta - \tfrac{\psi_n + \psi_{n'}}{2}\Bigr)\,\sin\!\Bigl(\tfrac{\psi_{n} - \psi_{n'}}{2}\Bigr),\\
C(\theta) \;&=\; 2R \,\cos(\theta) \,\cos\!\Bigl(\beta - \tfrac{\psi_n + \psi_{n'}}{2}\Bigr)\,\sin\!\Bigl(\tfrac{\psi_n - \psi_{n'}}{2}\Bigr).
\end{align}

\subsubsection{Correlation Coefficient Integral}
% In this paper, to be more suitable for the actual deployment of communication, we strictly consider only the case where the array radiates in half space. 
Under half-space isotropic scattering, the spatial correlation matrix of the arched URA can be formulated as
\begin{equation}
\scalebox{0.85}{$
\begin{aligned}
&[\mathbf{R}]_{(m,n),(m',n')} 
\\&= \int_0^\pi \int_0^\pi \frac{\sin(\theta)}{2\pi} \exp \left( j \frac{2\pi}{\lambda} \left[ r^{(m,n)} - r^{(m',n')} \right] \right) d\theta d\phi
\\&\approx \int_0^\pi \int_0^\pi \frac{\sin(\theta)}{2\pi} \exp \left( - j \frac{2\pi}{\lambda} \left[ \Delta_{m,n}(\theta,\phi) - \Delta_{m',n'}(\theta,\phi) \right] \right) d\phi\,d\theta
\\&\approx \int_0^\pi \int_0^\pi \frac{\sin(\theta)}{2\pi} \exp\!\Bigl(-j\,\tfrac{2\pi}{\lambda} \bigl[A(\theta)\cos(\phi) + B(\theta)\sin(\phi) + C(\theta)\bigr]\Bigr)\,d\phi\,d\theta.
\end{aligned}
$}
\label{URA1}
\end{equation}
\textit{{Solving the $\phi$-Integral:}}
For this complicated double integral, we first leverage the Bessel function identity~\cite{Ref_bowman2012introduction}
\begin{equation}
\int_0^\pi e^{-j\,\alpha \cos(\phi)}\,d\phi 
\;=\; 
\pi\,J_0(\alpha),
\end{equation}
then extend it to the linear combination $\alpha\cos(\phi) + \beta\sin(\phi)$
\begin{equation}
\scalebox{0.9}{$
\int_0^\pi \exp\!\Bigl(-j\,\tfrac{2\pi}{\lambda} \bigl(A\cos(\phi) + B\sin(\phi)\bigr)\Bigr)\,d\phi 
\;=\; 
\pi\, J_0\!\Bigl(\tfrac{2\pi}{\lambda}\,\sqrt{A^2 + B^2}\Bigr).
$}
\label{J2}
\end{equation}
By substituting \eqref{J2} into the spatial correlation expression \eqref{URA1}, we can obtain
\begin{equation}
\scalebox{0.9}{$
\begin{aligned}
&[\mathbf{R}]_{(m,n),(m',n')} \\&= \frac{1}{2} \int_0^\pi \sin(\theta) \, J_0\left( \frac{2\pi}{\lambda} \sqrt{A(\theta)^2 + B(\theta)^2} \right) \exp\left( - j \frac{2\pi}{\lambda} {C(\theta)}\right)\,d\theta.
\end{aligned}
$}
\label{eqURA}
\end{equation}
\textit{{Solving the $\theta$-Integral:}}
To better carry out the next step of integration, \eqref{eqURA} can be recast as
\begin{equation}
{\footnotesize
\begin{split}
[\mathbf{R}]_{(m,n),(m',n')} = \frac{1}{2} \int_0^\pi \sin(\theta) \, J_0\left( M D \sin(\theta) \right) e^{-j M E \cos(\theta)} \, d\theta,   
\end{split}
}
\end{equation}
where $M = \frac{2\pi}{\lambda}$ denotes the wavenumber. $D$ and $E$ are the identity transformation coefficients, where $D = \frac{\sqrt{A(\theta)^2 + B(\theta)^2}}{\sin(\theta)}$ and $E = \frac{{C(\theta)}}{\cos(\theta)}$.
% Then, we could easily substitute $x = \cos(\theta)$ in the remaining integral. Applying the integral identity
% \begin{equation}
% \int_{-1}^1 J_0\bigl(\kappa \sqrt{1-x^2}\bigr)\,e^{\,j\,\eta x}\,dx 
% \;=\; 
% \frac{2\,\sin\!\bigl(\sqrt{\kappa^2 + \eta^2}\bigr)}{\sqrt{\kappa^2 + \eta^2}},
% \end{equation}
% \textbf{Theorem 2.} The closed-form spatial correlation solution of arched URA can be written as
% \begin{equation}
% \bigl[\textbf{R}\bigr]_{(m,n),(m',n')} 
% \;=\; 
% \frac{\lambda}{2\pi\,\sqrt{D^2 + E^2}}\, \sin\!\Bigl(\frac{2\pi}{\lambda}\,\sqrt{D^2 + E^2}\Bigr),
% \label{eq:Rmn_final}
% \end{equation}
% where
% \begin{equation}
% \begin{cases}
% \scalebox{0.86}{$
% \begin{aligned}
% D \;&=\; \sqrt{\,(m - m')^2\,d^2 \;+\; 4\,R^2\,\sin^2\Bigl(\tfrac{\psi_n - \psi_{n'}}{2}\Bigr)\,\cos^2\Bigl(\beta - \tfrac{\psi_n + \psi_{n'}}{2}\Bigr)},\\[6pt]
% E \;&=\; 2\,R \,\cos\!\Bigl(\beta - \tfrac{\psi_n + \psi_{n'}}{2}\Bigr)\,\sin\!\Bigl(\tfrac{\psi_n - \psi_{n'}}{2}\Bigr).
% \end{aligned}
% $}
% \end{cases}
% \end{equation}

Finally, by setting \( x = \cos(\theta) \) in the remaining integral and applying the following integral identity~\cite{Ref_bowman2012introduction},
\begin{equation}
\int_{-1}^1 J_0\bigl(\kappa \sqrt{1-x^2}\bigr)\,e^{\,j\,\eta x}\,dx 
\;=\;
\frac{2\,\sin\!\bigl(\sqrt{\kappa^2 + \eta^2}\bigr)}{\sqrt{\kappa^2 + \eta^2}},
\end{equation}
we arrive at the closed-form expression for the spatial correlation of the arched URA. The result is summarized in Theorem 2 below.
\\
\textbf{Theorem 2.} \textit{The closed-form spatial correlation for the arched URA can be written as}
\begin{equation}
\bigl[\mathbf{R}\bigr]_{(m,n),(m',n')} 
\;=\; 
\frac{\lambda}{2\pi\,\sqrt{D^2 + E^2}}\,
\sin\!\Bigl(\frac{2\pi}{\lambda}\,\sqrt{D^2 + E^2}\Bigr),
\label{eq:Rmn_final}
\end{equation}
\textit{where}
\begin{equation}
\begin{cases}
\scalebox{0.86}{$
\begin{aligned}
D \;&=\; \sqrt{\,(m - m')^2\,d^2 \;+\; 4\,R^2\,\sin^2\Bigl(\tfrac{\psi_n - \psi_{n'}}{2}\Bigr)\,\sin^2\Bigl(\beta - \tfrac{\psi_n + \psi_{n'}}{2}\Bigr)},\\[6pt]
E \;&=\; 2\,R \,\cos\!\Bigl(\beta - \tfrac{\psi_n + \psi_{n'}}{2}\Bigr)\,\sin\!\Bigl(\tfrac{\psi_n - \psi_{n'}}{2}\Bigr).
\end{aligned}
$}
\end{cases}
\end{equation}
% \textbf{Remark 2.} 
% From \textbf{Theorem 2}  we can enable a straightforward analysis of spatial correlation and degrees of freedom across various arched URA geometries.
% (including URA and semi-cylindrical arrays, which is further demonstrated below).
\subsubsection{Special Cases}
Now, let us discuss two limiting cases.
\\
\textit{Case 1: $\beta = 0$ (URA Limit).}
When $\beta = 0$, $R \to \infty$ and with small $\psi_n$,
% we can use $\sin x \approx x$ and $\cos x \approx 1$.
the expression is simplified to
% \begin{align}
% D &= \sqrt{(m - m')^2 d^2 \;+\; 4R^2 \,\sin^2\Bigl(\tfrac{\psi_n - \psi_{n'}}{2}\Bigr)\,\sin^2\Bigl(\tfrac{\psi_n + \psi_{n'}}{2}\Bigr)},
% \nonumber \\
% E &= 0.
% \end{align}
\begin{equation}
% D \;\approx\; |m - m'|\,d
% \quad\Longrightarrow\quad
\bigl[\mathbf{R}\bigr]_{(m,n),(m',n')} 
\;=\; 
\frac{\lambda}{2\pi \,|m - m'|\,d} 
\,\sin\!\Bigl(\tfrac{2\pi}{\lambda}\,|m - m'|\,d\Bigr),
\end{equation}
which is in good agreement with the traditional URA correlation~\cite{Ref_SunTao2022,newpizzo2020spatially,bjornson2020rayleigh}.
\\
\textit{Case 2: $\beta = \pi/2$ (Semi-Cylinder).}
When $\beta = \pi/2$, we obtain
\begin{align}
D &= \sqrt{ (m - m')^2 d^2 + 4R^2 \sin^2\Bigl(\tfrac{\psi_n - \psi_{n'}}{2}\Bigr)\,\cos^2\Bigl( \tfrac{\psi_n + \psi_{n'}}{2}\Bigr)}, \label{eq:D} \\
E &= R (\cos\psi_{n'} - \cos\psi_n), \label{eq:E}
\end{align}
where \( R = \frac{L}{\pi} \) and \( \psi_n = \frac{(n-1)\pi}{N-1} \), the result aligns with the conventional semi-cylindrical configuration.\footnote{The closed-form solution for the spatial correlation of the semi-cylinder, which corroborates the findings presented here, is not demonstrated in this paper due to space limitations.}

Through verification, we find that this closed-form solution unifies the planar and curved geometries in a single framework, which not only facilitates efficient numerical evaluations but also provides clear insights into how curvature impact spatial correlation across the array surface.

\section{Simulation Results}

In this section, we validate the theoretical analysis through numerical simulations. Highlighting the impact of curvature on spatial correlation and DoF. We select a practically achievable and small array size, where the array element spacing is one-fifth of the wavelength, to observe the simulation performance under arched HMIMO. We consider the following configurations:

\begin{itemize}
    \item \textbf{Arched ULA:} \(N = 512\) elements operating at \(f = 100\ \mathrm{GHz}\) (\(\lambda = 3\ \mathrm{mm}\)), total arc length \(L = 0.3142\ \mathrm{m}\), element spacing \(d = L/(N-1)\).
   \item \textbf{Arched URA:} \(N = 64*64\), physical length \(L_x = L_z = L = 0.0393\ \mathrm{m}\), curvature parameter \(\beta \in \{0, \pi/2\}\). 
\end{itemize}
We observe several key phenomena that can be explained by the structure of the closed-form correlation expression. 
Define $\alpha_n=\frac{n-1}{N-1}\cdot\frac{L}{R}$, and $R=\frac{L}{2\beta}$ is the curvature radius. Substituting these definitions into (26), the correlation formula of arched ULAs becomes
\begin{equation}
[\mathbf{R}]_{m,n} = \text{sinc}\left(\frac{2L}{\lambda\beta}\sin\left(\frac{n-m}{N-1}\beta\right)\right). 
\end{equation}
The correlation thus explicitly depends on the element angular separation. 
% and the largest argument (at maximum element spacing $|n-m|=N-1$) is $\frac{2L}{\lambda\beta}\sin\beta$.

For arched URAs, spatial correlation is similarly characterized using a sinc function dependent on inter-element distance, by transforming formula (39)
\begin{equation}
[\mathbf{R}]_{(m,n),(m',n')}=\text{sinc}\left(\frac{2\pi}{\lambda} d_{(m,n),(m',n')}\right),
\end{equation}
the distance between elements is
\begin{equation}
\begin{aligned}
d_{(m,n),(m',n')}^2&=(m-m')^2 d_x^2+4R^2\sin^2\left(\frac{(n-n')\beta}{N-1}\right),\\&R=\frac{L}{2\beta},\quad d_x=\frac{L}{M-1}.
\end{aligned}
\end{equation}
As $\beta\to 0$, (45) naturally reduces to the classical planar URA expression based on Euclidean distance, demonstrating its generality.

Fig. 3 and Fig. 4 illustrate the eigenvalue distributions of the arched ULA and arched URA, respectively. It can be observed from the two figures that both the arched ULA and URA each approximately maintain their respective counts of dominant eigenvalues, thereby exhibiting individually nearly constant DoF across the entire curvature range.

However, the distinctions in correlation structures are notable. Arched ULAs are inherently one-dimensional structures, with spatial correlation governed exclusively by angular separation. Conversely, arched URAs, as two-dimensional arrays, introduce complexity via distances in both horizontal and curved vertical dimensions. Consequently, the impact of curvature is more nuanced in URAs, while horizontal spacing remains constant, vertical curvature may compress the effective aperture, resulting in modest reductions in DoF.

\begin{figure}[t]
\centerline{\includegraphics[width=1\linewidth]{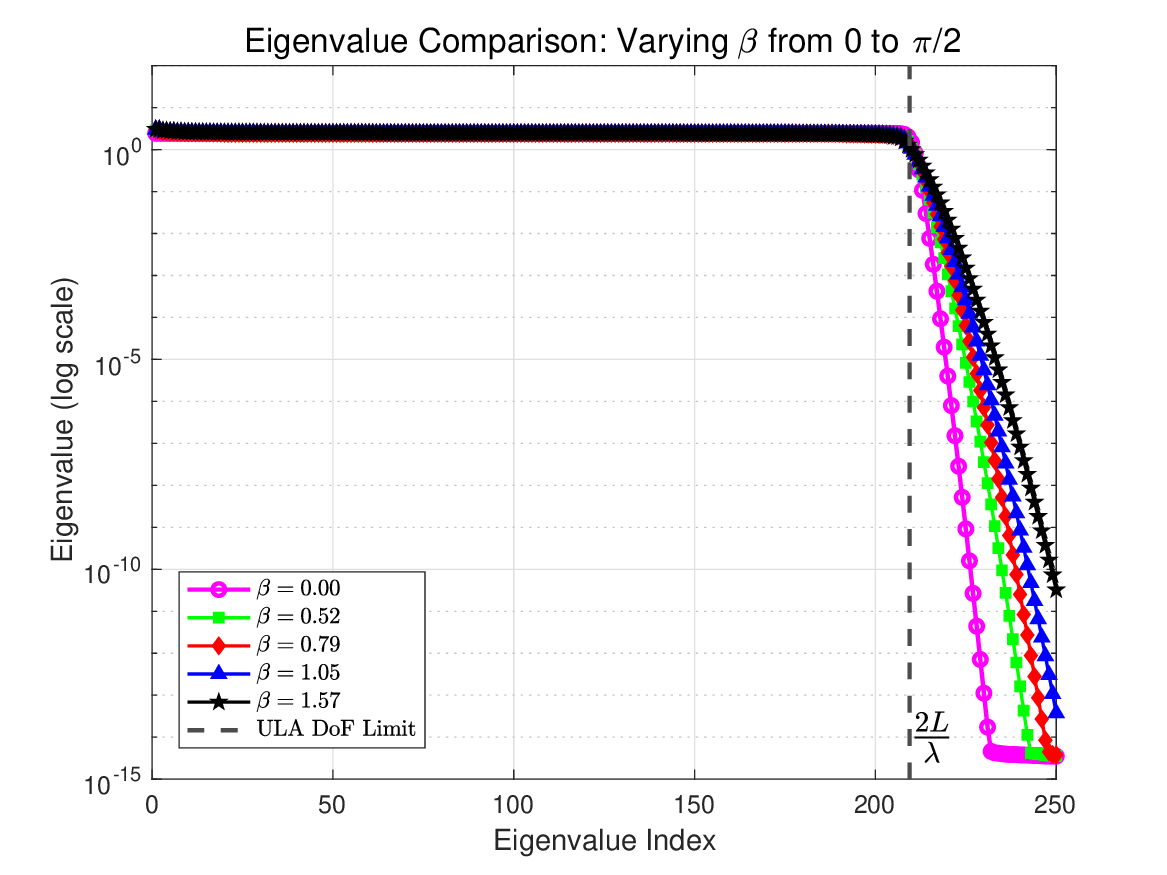}}
\captionsetup{font=footnotesize, labelsep=period}
\caption{Eigenvalue versus eigenvalue index of spatial correlation of arched ULA in decreasing order for various element spacing $d_{yz}$ and with $L = 0.3142\ \mathrm{m}$. Additionally, the asymptotic spatial degrees of freedom of the ULA $\frac{2L}{\lambda}$ are plotted. Note that only the array length is fixed to $\L$, and the eigenvalues and their indices are dimensionless.}

\label{ULA4}
\end{figure}
\vspace{2ex} 
\begin{figure}[t]
\centerline{\includegraphics[width=1\linewidth]{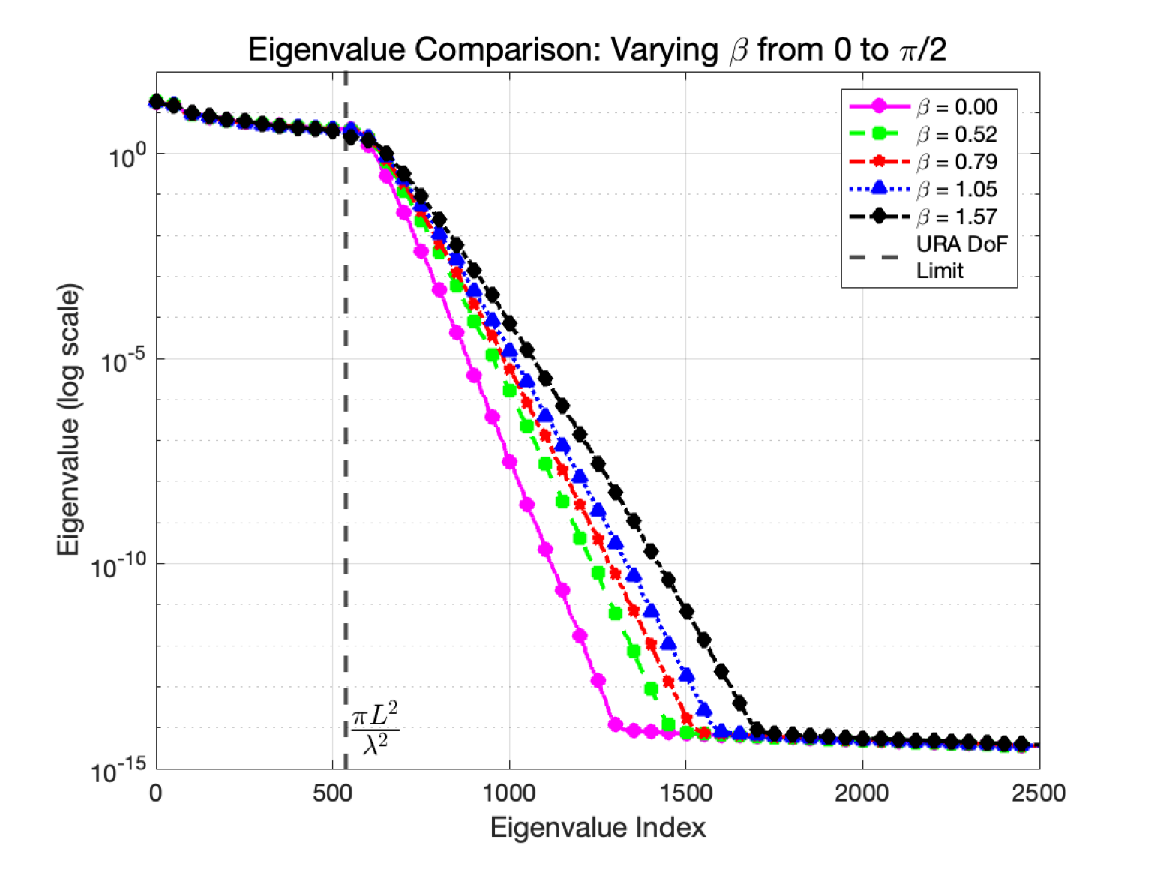}}
\captionsetup{font=footnotesize, labelsep=period}
\caption{Eigenvalue versus eigenvalue index of spatial correlation of arched URA in decreasing order for various element spacing $d_{yz}$, $d_{x}$ and with $L = 0.0393\ \mathrm{m}$. In addition, the asymptotic spatial degrees of freedom $\frac{\pi L^2}{\lambda^2}$ derived in~\cite{Ref_PizzoMarzetta2020} is drawn in the figure. Both the eigenvalue and its index are dimensionless.}
\label{URA2}
\end{figure}
% Both arched ULA and URA results align with the closed-form solutions in Theorems 1–2, bridging planar and curved geometries. 
% Notably, the arched URA’s non-monotonic DoF behavior challenges conventional design paradigms, while the arched ULA’s robustness highlights the practical viability for reconfigurable HMIMO surfaces. These insights enable geometry-aware array optimization, balancing curvature for desired diversity-resolution tradeoffs. 

\section{Conclusion}

In this paper, we presented a comprehensive closed-form analysis of the spatial correlation and DoF for arched HMIMO arrays, considering both arched ULAs and URAs in 3D space. Inspired by the fact that real-world HMIMO surfaces may be curved or fluidically adaptable as in FAS, we derived key analytical expressions for the spatial correlation, demonstrating how curvature influences overall performance.

For arched HMIMO (both the arched ULAs and URAs), the DoF exhibits approximately constant behavior with increasing curvature. This indicates a high degree of robustness to array bending, even under significant geometric deformation. In particular, the DoF remains relatively stable across a wide range of curvatures, suggesting that moderately curved array implementations may not significantly compromise spatial performance.

%The closed-form results for both geometries provide useful insights for designing practical HMIMO arrays, where curved surfaces or conformal installations are likely in real-world scenarios. Our analysis supports the development of flexible and efficient HMIMO systems that maintain strong spatial diversity and resolution even in the presence of curvature.
In summary, this work bridges planar and curved HMIMO geometries (viewable as a type of FAS whenever geometry is reconfigurable), offering a unified framework for quantifying spatial correlation and DoF limits. The findings emphasize that curvature-aware design enables geometric adaptability without substantial performance loss, paving the way for resilient HMIMO implementations in complex environments.

\bstctlcite{IEEEexample:BSTcontrol}
\bibliographystyle{IEEEtran}
\bibliography{2025VTC}

\end{document}